\documentclass[twocolumn,showpacs,preprintnumbers,amsmath,amssymb]{revtex4}
\usepackage{graphicx}% Include figure files
\usepackage{dcolumn}% Align table columns on decimal point
\usepackage{bm}% bold math

\usepackage{amsmath}
\usepackage{amssymb}
\usepackage{subfigure}
\usepackage{feynmf}
\usepackage{natbib}
\newcommand{\Tau}{\mathcal{T}}

\begin{document}

%\preprint{APS/123-QED}

\title{Quantum Multicriticality}

\author{G.T. Oliver}
\author{A.J. Schofield}
\affiliation{%
School of Physics and Astronomy, University of Birmingham, Edgbaston, Birmingham, B15 2TT, 
United Kingdom.
}%

\date{\today}% It is always \today, today,

\begin{abstract}
Several quantum critical compounds have been argued to have multiple instabilities towards orders with distinct dynamical exponents. We present an analysis of a quantum multicritical point in an itinerant magnet with competition between ferro- and antiferromagnetic order, modelled using Hertz-Millis theory. We perform a one-loop renormalization group treatment of this action in the presence of two dynamical exponents. In two and in three dimensions, when both incipient orders are quantum critical, we find that the specific heat, thermal expansion and Gr\"{u}neisen parameter obey the same power laws as those expected for a single ferromagnetic quantum critical point. The antiferromagnetic correlation length and boundary of the antiferromagnetic ordered phase are suppressed by the dangerously irrelevant interactions with quantum critical ferromagnetic fluctuations. We find no difference between a quantum bicritical point and a quantum tetracritical point. Our results are compared with experiments on NbFe$_2$.
\end{abstract}

\pacs{73.43.Nq, 71.10.Hf, 64.40.Kw}
\maketitle

%\section{Introduction}
Quantum criticality is characterised by universal divergences of thermodynamic quantities at a continuous zero-temperature phase-transition as some non-thermal control parameter (e.g. pressure, doping or magnetic field) is changed. The much-studied power laws associated with the quantum phase transition depend on the universality class, which in contrast to the classical case depends on the dynamics of the order parameter fluctuations in imaginary time. The dynamic effects are characterized by a dynamical exponent, $z$, which, in addition to controlling the power-law divergences, defines the boundaries of distinct regions in the phase diagram \cite{qpts,lrv07,h76,m93}.

In recent years quantum critical behaviour has been observed in systems that do not seem to conform to the standard theoretical picture of fluctuations of an order parameter with a unique dynamical exponent. For example the specific heat in NbFe$_2$ displays a $C \sim -T\ln T$ relation as would be expected for three dimensional itinerant ferromagnetic quantum criticality which is described with $z=3$. In contrast the resistivity displays $\rho \sim T^{3/2}$ as expected for a dirty three dimensional antiferromagnet, usually described with $z=2$ \cite{mba09,bma08}. 

In this Letter we report results of a study into quantum criticality where an itinerant material is unstable to both ferro- and antiferromagnetic order. Materials with a multicritical point in the phase diagram [as shown in Figs. 1(a) and 1(b)] show a competition between two distinct types of order. Here we assume that if we had control over another tuning parameter we could suppress this multicritical point to zero temperature, to form a quantum multicritical point [as shown in Fig. 1(c)]. This type of phase diagram has been considered before \cite{tgg13}, but in the presence of a symmetry-breaking field, which we do not treat here. To model quantum multicriticality we construct an effective action in terms of spin-fluctuations by adapting Hertz-Millis theory \cite{h76,m93,gthesis}. This neglects important non-analytic terms which sometimes arise from integrating out the fermionic degrees of freedom \cite{lrv07}. Ignoring such terms seems valid slightly away from the critical point \cite{smith_2008a}. Nevertheless
their inclusion could stabilize a finite-momentum spin-density wave near the
ferromagnetic quantum critical point~\cite{conduit} to generate the scenario considered
here. We extend Millis' calculation \cite{m93} to treat the quantum multicritical point by following the renormalization group (RG) procedure. This enables us to map out the rich phase diagram and calculate the leading order critical parts of the specific heat, thermal expansion and Gr\"{u}neisen parameter.

Our main result is that, since we are at or above the upper critical dimension of our model ($d_c^+=2$), we can essentially treat the fluctuations of each type of order independently, and the critical part of the free energy and therefore its derivatives simply become the sum of the contributions associated with each individual order parameter. The caveat is that whenever the \emph{dangerously irrelevant} interactions affect the RG flow, due to the multiple dynamics they can produce novel temperature dependences. This allows the dangerously irrelevant interactions to shape the phase diagram. We also discuss the resistivity in relation to experiments, though do not offer an explicit calculation here.
\begin{figure}
 \begin{center}
 \begin{tabular}{c c}
 \subfigure[]{\includegraphics[scale=0.3]{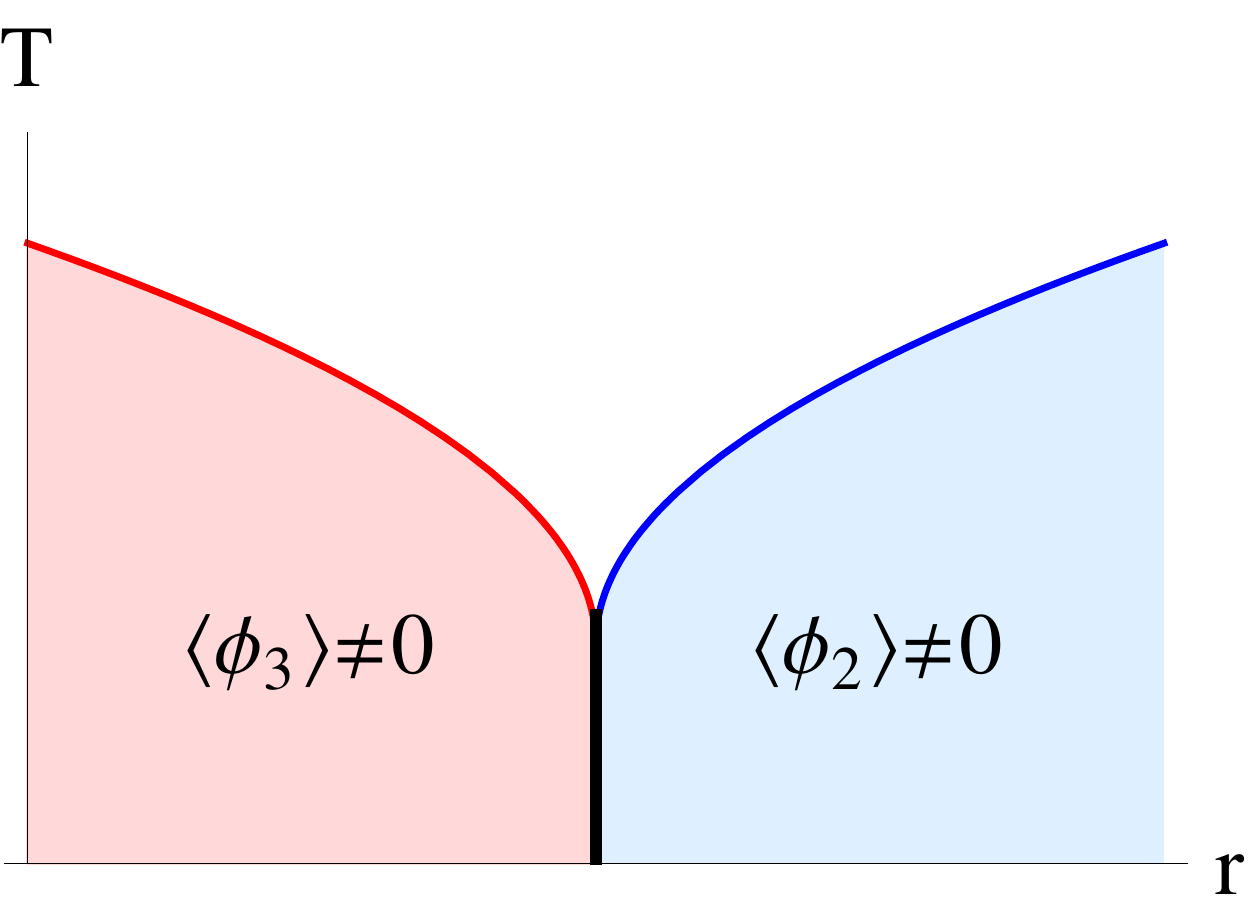}\label{fig:subfig1}} & 
 \subfigure[]{\includegraphics[scale=0.3]{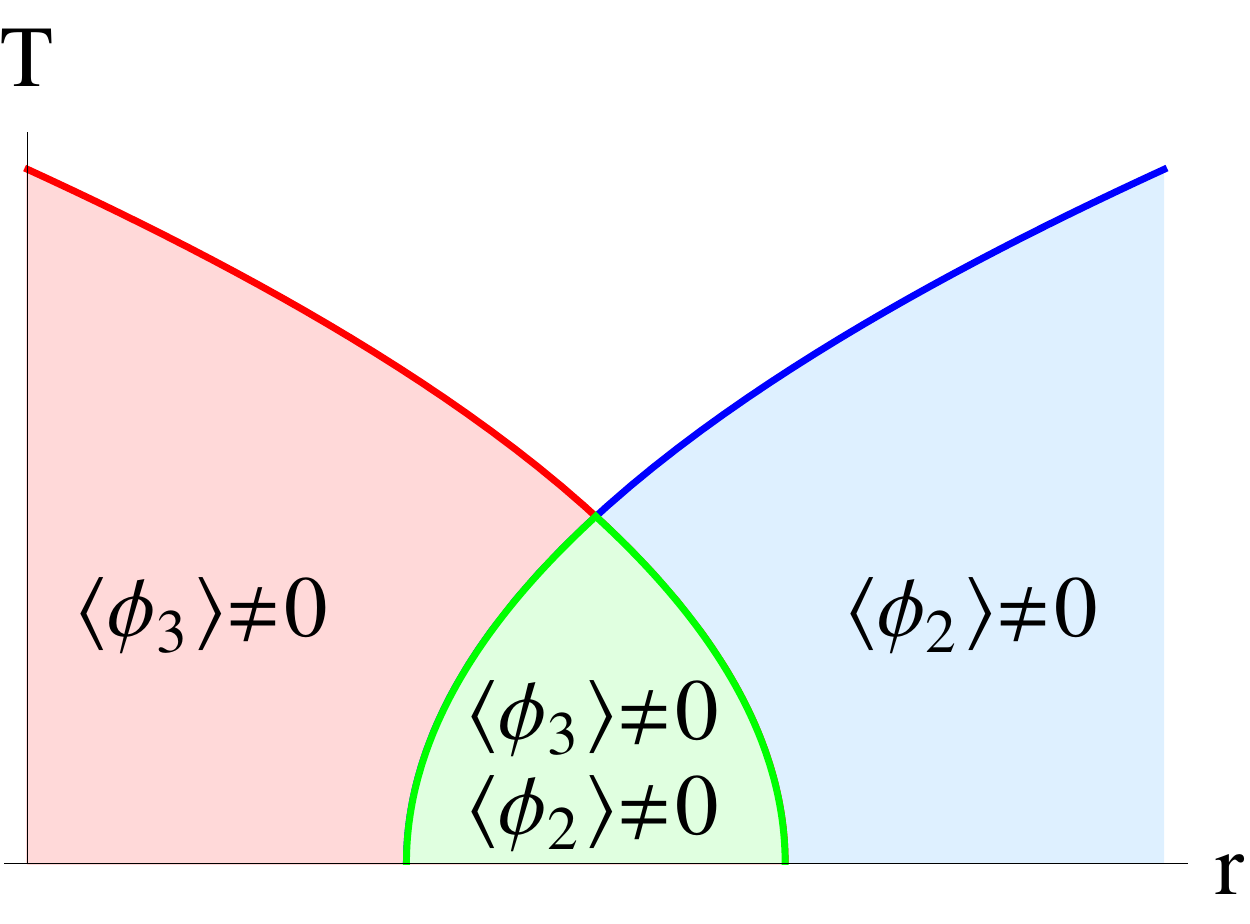}\label{fig:subfig2}}
 \end{tabular}\\
 \end{center}
  \begin{center}
 \begin{tabular}{c c}
 \subfigure[]{\includegraphics[scale=0.3]{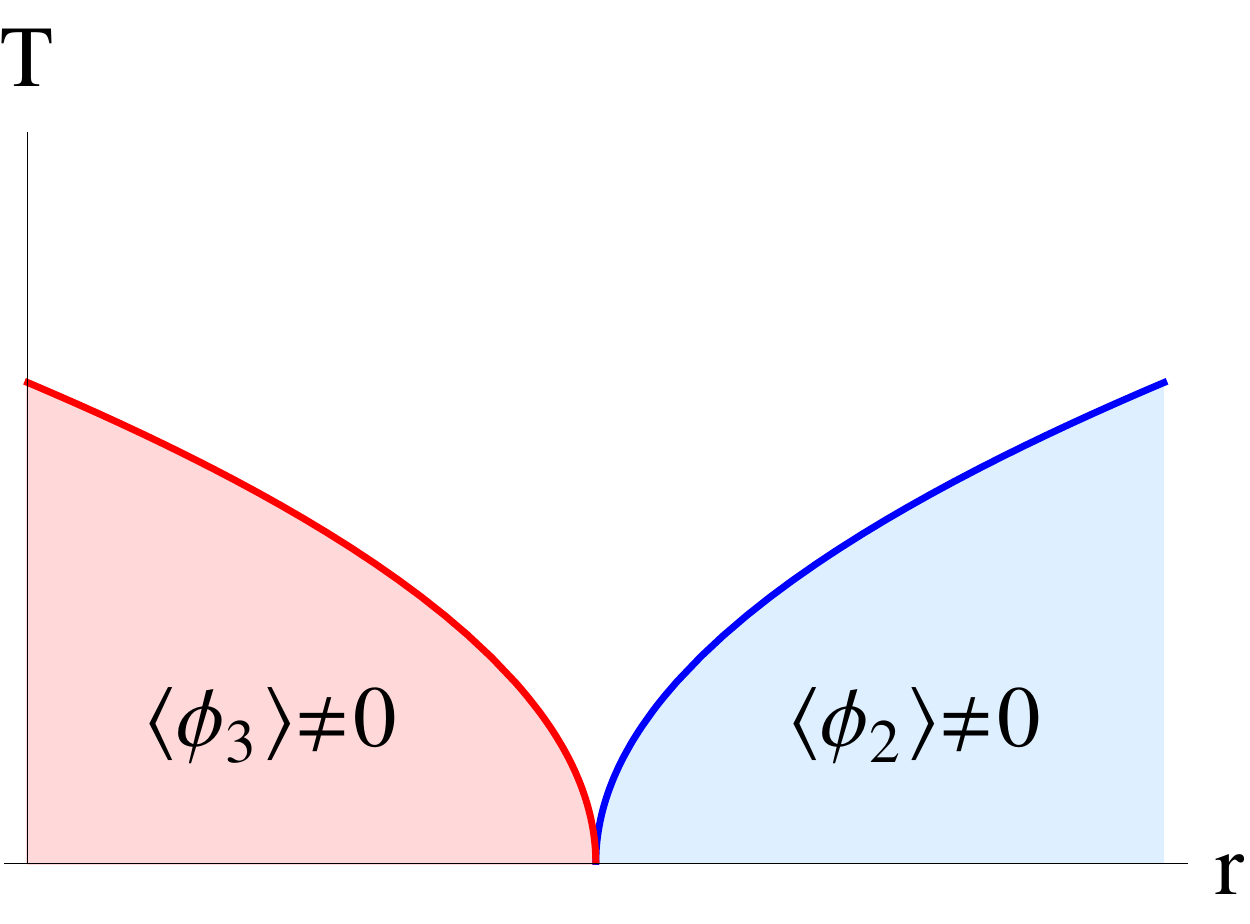}\label{fig:subfig3}} & 
 \subfigure[]{\includegraphics[scale=0.3]{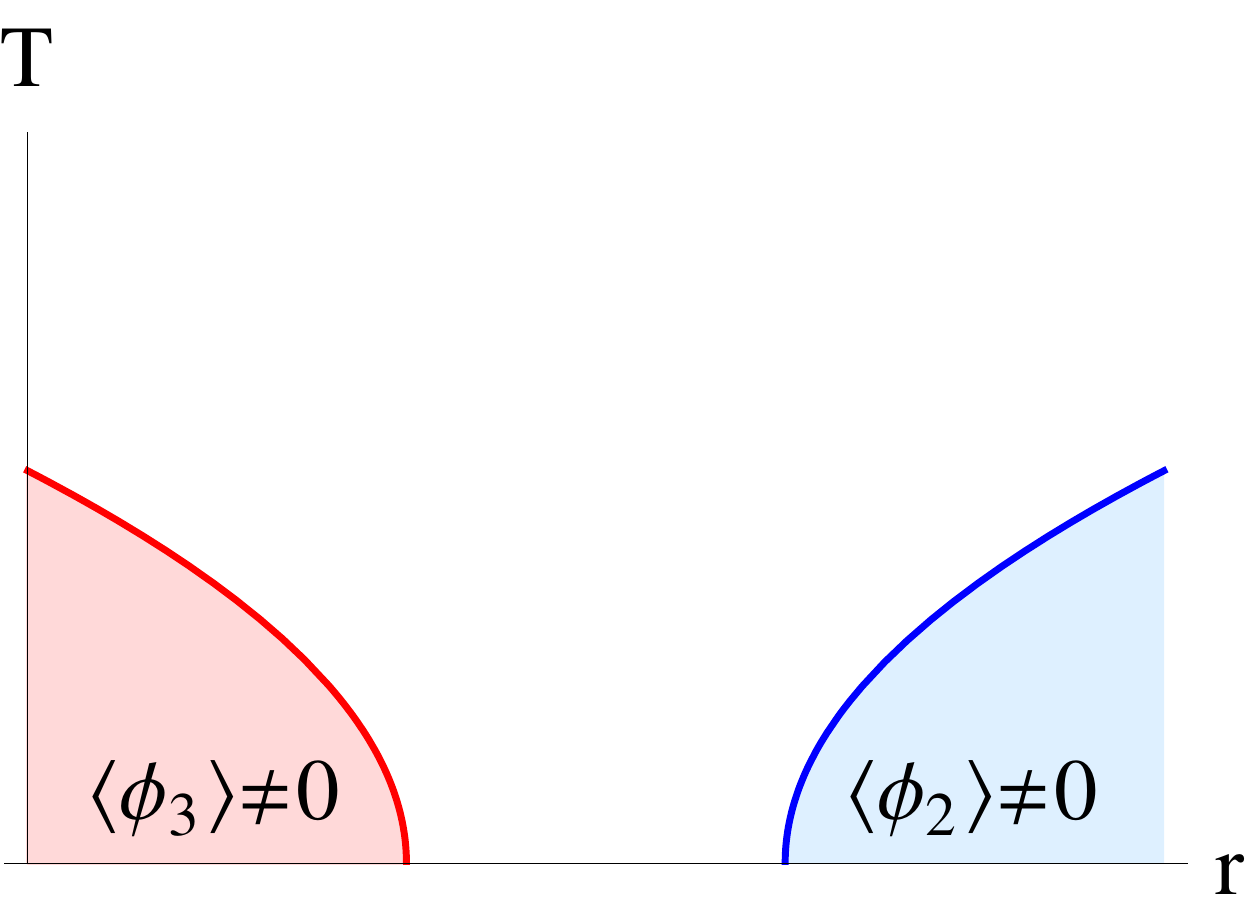}\label{fig:subfig4}}
 \end{tabular}
 \end{center}
\caption{Suppressing a multicritical point to zero temperature. (a) and (b) show bicritical and tetracritical points respectively. By invoking another non-thermal control parameter $g$, these multicritical points can be tuned to zero temperature, as shown in (c). (d) shows the expected phase diagram when two quantum critical points arise for increasing $g$.
}
\end{figure}

%\section{Quantum Multicritical Point Action}
Our analysis is formulated in terms of an order parameter field $\phi(\bm{q},\omega_n)$ which describes the magnetization of the system. In a system unstable towards both ferro- and antiferromagnetic order, the susceptibility will be large near zero momentum and near the antiferromagnetic wavevector $\bm{Q}$. We split the magnetization field into two parts to represent these different regimes. We denote the small momentum part of the field as $\phi_3(\bm{q},\omega_n)$ for $|\bm{q}|<\Lambda_3$ where the subscript $3$ refers to the fact that this field is usually described by a dynamical exponent $z=3$. To denote the field near momentum $\bm{Q}$ we use the notation $\phi_2(\bm{q},\omega_n)$ where here $\bm{q}$ (which we restrict such that $|\bm{q}|<\Lambda_2$) measures the deviation from $\bm{Q}$. Here the subscript $2$ refers to the dynamical exponent $z=2$ usually associated with antiferromagnetic order. We allow for $\phi_3$ and $\phi_2$ to have $n_3$ and $n_2$ components respectively.

The action we use to describe a quantum multicritical point is the sum of the actions of a ferromagnetic and an antiferromagnetic quantum critical point (QCP),
\begin{multline}
S\left[\phi_3,\phi_2\right] = \sum_{i=3,2}\sum_{\omega_n}\sum_q
\chi_i^{-1}\left(\bm{q},\omega_n\right)\phi_i^2\left(\bm{q},\omega_n\right)\\
+ \int d\bm{x}d\tau
\left[ u_3 \phi_3^4(\bm{x},\tau) + u_2\phi_2^4(\bm{x},\tau)\right.\\
+ \left.u_{32}\phi_3^2(\bm{x},\tau)\phi_2^2(\bm{x},\tau) \right],
\end{multline}
where the two QCPs are coupled together by a mode-mode coupling term $u_{32}$. The bare inverse spin susceptibilities are given by
\begin{eqnarray}
\chi_i^{-1}(\bm{q},\omega_n) = \delta_i + q^2 + \eta_i\frac{|\omega_n|}{q^{{z_i}-2}},
\end{eqnarray}
where $z_3$ is the dynamical exponent associated with ferromagnetic order, $z=3$, and $z_2$ is the dynamical exponent associated with antiferromagnetic order, $z=2$. We have added `kinetic coefficients' $\eta_3$ and $\eta_2$ which allow us to renormalize in the imaginary time direction \cite{zwg09,mrg12}. The classical analogue of the action describes a multicritical point in the $\delta_3$-$\delta_2$ plane. The model shows bicriticality if $u_{32}^2 > 4u_3u_2$ and tetracriticality if this is inequality is reversed \cite{knf76,pocmp}.

%\section{Renormalization Group Analysis}
In order to map out the phase diagram and predict thermodynamic quantities, we perform an RG analysis by simultaneously integrating out the $\phi_3$ modes in a small shell with momenta between $\Lambda_3/b$ and $\Lambda_3$, and the $\phi_2$ modes with momenta between $\Lambda_2/b$ and $\Lambda_2$. We then rescale such that the original cut-offs are restored and calculate how the other parameters in the model must rescale. The presence of multiple dynamical exponents means there is no unique way to rescale frequency. We choose to rescale it as $b^z$ where $z$ is a fictitious dynamical exponent which we leave unspecific (see Ref. \cite{zwg09} and \cite{mrg12}). This enables renormalization but will drop out of our calculations so that no physical properties depend on it. The RG equations can either be derived directly from the action or by calculating a physical property (such as the free energy) and ensuring it does not change under renormalization. We have done both, and find that the one-loop RG equations for the tuning parameters and interactions are
\begin{subequations}
\begin{multline}
\frac{d\delta_i}{d\ln b} = 2\delta_i(b) + \tilde{u}_i(b) 4(n_i+2)f_i^{(2)}\left(\delta_i(b),\Tau_i(b)\right)\\
+ \tilde{w}_{\bar{\imath}}(b)4n_{\bar{\imath}}f_{\bar{\imath}}^{(2)}\left(\delta_{\bar{\imath}}(b),\Tau_{\bar{\imath}}(b)\right),
\end{multline}
\begin{multline}
\frac{d\tilde{u}_i}{d\ln b} = \left[4-(d+z_i)\right]\tilde{u}_i(b) - 4(n_i+8)f_i^{(4)}(\delta_i(b),\Tau_i(b))\tilde{u}_i^2\\
 - 4n_{\bar{\imath}}f_{\bar{\imath}}^{(4)}(\delta_{\bar{\imath}}(b),\Tau_{\bar{\imath}}(b))\tilde{w}_i(b)\tilde{w}_{\bar{\imath}}(b),
\end{multline}
\begin{multline}
\frac{d\tilde{w}_i}{d\ln b} = \left[4-(d+z_i)\right]\tilde{w}_i(b)\\
- \sum_{j=3,2} 4\left(n_j+2\right)f_j^{(4)}(\delta_j(b),\Tau_j(b))\tilde{u}_j(b)\tilde{w}_i(b),
\end{multline}
\end{subequations}
where $i$ is either 3 or 2 and $\bar{\imath}$ is correspondingly either 2 or 3. We have defined $\tilde{u}_i\equiv u_i/\eta_i$ and $\tilde{w}_i\equiv u_{32}/\eta_i$, as we find that the only way the interactions enter the RG equations is in these combinations. For the rest of this Letter, when a parameter is written without explicit scale-dependence we are referring to the bare, unrenormalized value. We find that the temperature scales as $T(b)=b^zT$ and the kinetic coefficients are $\eta_i(b) = b^{z_i-z}\eta_i$. The RG equations can be written in terms of two temperature fields $\Tau_i(b)= \eta_i(b)T(b) = \Tau_i b^{z_i}$ which represent the effective temperature felt by the $\phi_i$ modes. The $f^{(2)}_i$ and $f^{(4)}_i$ functions are the one-loop integrals shown in Fig. 2, which arise in the RG procedure due to interactions with modes above the cut-off. They are exactly the same functions which appear in Hertz-Millis theory, defined in Ref. \cite{m93}.
\begin{figure}
\begin{center}
\begin{tabular}{c c}
\subfigure[]{\includegraphics[scale=0.44]{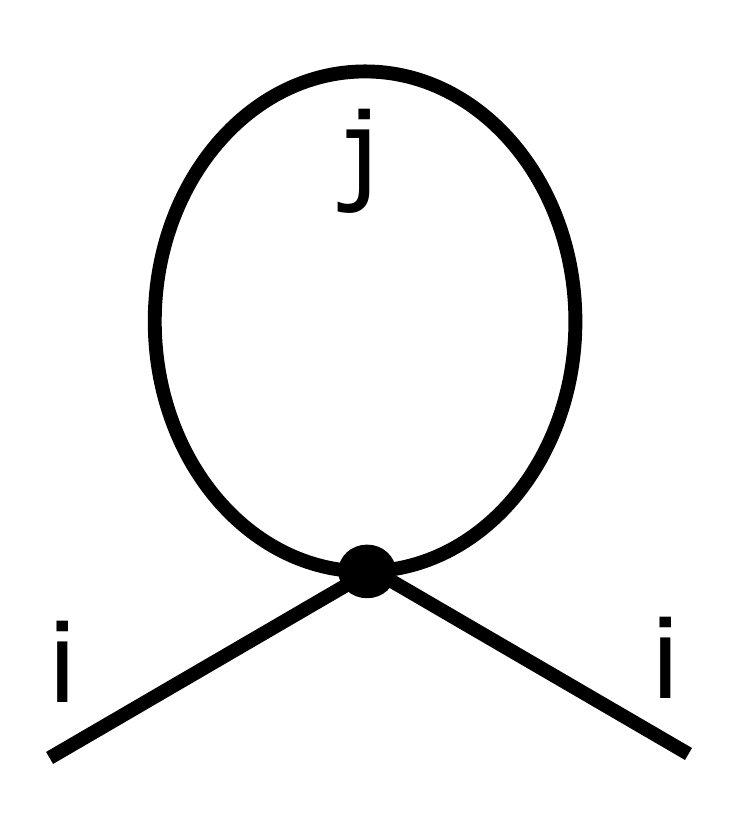}\label{fig:subfig1}} &
\subfigure[]{\includegraphics[scale=0.44]{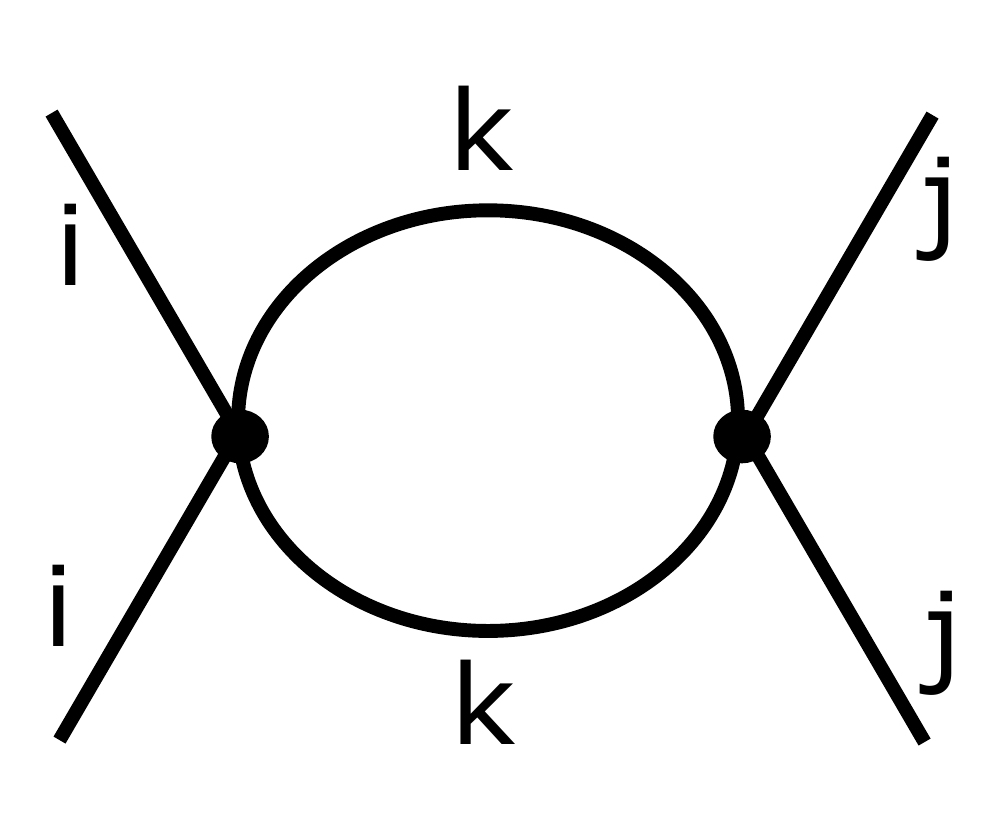}\label{fig:subfig2}}
\end{tabular}
\caption{One-loop diagrams that contribute to the RG flow and Eqs. (3a), (3b) and (3b). Lines with free (connected) ends represent modes below (above) the cut-off. The symbols $i,j,k=3,2$ label which modes the lines represent. (a) shows a contribution to the renormalization of $\delta_i$ from interaction with $\phi_j$ modes above the cut-off, and is proportional to $f^{(2)}_j\left(\delta_j,\Tau_j\right)$. (b) shows a contribution to the renormalization of $u_{i}$ if $j=i$ and $u_{32}$ if $j\neq i$, due to interactions with $k$ modes above the cut-off, and is proportional to $f^{(4)}_k\left(\delta_k,\Tau_k\right)$.}
\end{center}
\end{figure}
%\subsection{Solution}
The scaling of the interaction terms can be calculated from Eqs. (3b) and (3c), which can be analysed under the usual approximation that the $f^{(4)}$ functions are constant \cite{m93}. In $d=3$ we find all interaction terms decay as some power of $b$, whereas in $d=2$, $\tilde{u}_2$ and $\tilde{w}_2$ decay logarithmically. We find that at large values of $b$, $\tilde{u}_2(b) \sim \left(\ln b\right)^{-1}$ just as in the $d=z=2$ Hertz-Millis case, and $\tilde{w}_2(b) \sim \left(\ln b\right)^{-(n_2+2)/(n_2+8)}$, which is a very slow decay unique to the multicritical case. Because the free energy can be written as a power series in these interaction terms, we conclude that the upper critical dimension is $d_c^+=2$, just as for an antiferromagnetic QCP. In the cases of two and three dimensions considered here, we are at or above the upper critical dimension and so are controlled by the Gaussian fixed point where all interactions flow to zero.

The distinct regions of the phase diagrams and the correlation lengths in each regime can be calculated from Eq. 3(a). In our calculation we use Millis' approximation for the integral of the $f^{(2)}(\delta,\Tau)$ function, which is different in the quantum critical and quantum disordered regimes \cite{m93}, defined by $R \ll \Tau^{2/z}$ and $R \gg \Tau^{2/z}$ for a single QCP. Here $R$ is the renormalized tuning parameter or quasiparticle mass, which may acquire some temperature dependence. In the multicritical case, we conclude that both the ferro- and antiferromagnetic modes can independently be quantum critical or quantum disordered, splitting the phase diagram into four regions separated by the two lines $R_i \sim \Tau_i^{2/z_i}$. The solution of Eq. 3(a) at large values of $b$ yields a tuning parameter becomes $\delta_i(b)=b^2R_i$ where $R_i$ is renormalized by interactions with both modes independently. This can be related to the correlation length of the corresponding order parameter by $R_i = \xi_i^{-2}$. We denote the zero temperature part of $R_i$ by $r_i$, and it is this which tunes to the QCP at $r_i=0$.

It is the dangerously irrelevant interaction terms which give the renormalized tuning parameters their temperature dependence, which in turn control the correlation lengths and the boundaries of the ordered phases. In three dimensions the boundaries of the ordered phases can be calculated from the lines $T(r_i)$ where the corresponding correlation length diverges. While no true order can exist in 2D, we adopt the usual convention and use the point that the Ginzburg criterion of the classical theory breaks down to identify the `phase boundary'.

The generic phase diagram for a quantum multicritical point is shown in Fig. 3, which is qualitatively the same in both two and three dimensions. We find it most revealing to interpret the results as the sum of two quantum critical points. In both two and three dimensions, we find that to leading order the ferromagnetic $z=3$ QCP is qualitatively unaffected by the antiferromagnetic $z=2$ QCP. However the antiferromagnetic QCP is strongly affected by the proximity to a ferromagnetic QCP. When the ferromagnetic modes are quantum critical, the antiferromagnetic correlation length acquires the temperature dependence of the ferromagnetic correlation $\xi_2^{-2}\sim r_2 + AT^{4/3}$ instead of the usual $\xi_2^{-2}\sim r_2+BT^{3/2}$ in three dimensions, and $\xi_2^{-2}\sim r_2+ CT\ln\left(1/T\right)$ instead of the usual $\xi_2^{-2} = r_2 + D\ln\left(\ln\left(1/T\right)\right)/\ln\left(1/T\right)$ in two dimensions. This temperature dependence dominates the antiferromagnetic correlation length in region I of the phase diagram in Fig. 3. Interactions with quantum critical ferromagnetic fluctuations therefore reduce the antiferromagnetic correlation length, which in turn suppresses the boundary of the antiferromagnetic phase. If the ferromagnetic fluctuations are Fermi liquid-like then the antiferromagnetic QCP is qualitatively unaffected.

Since the interactions are irrelevant, thermodynamic properties in the disordered region of the phase diagram can be obtained from the Gaussian part of the free energy, which can be calculated directly from the action. This is just the sum of the contributions from both individual QCPs weighted by $\eta_i^{-1}$, $F_G(T,\delta_3,\delta_2) = \frac{1}{\eta_3}F_G^{(3)}\left(\delta_3,\Tau_3\right)+\frac{1}{\eta_2}F_G^{(2)}\left(\delta_2,\Tau_2\right)$, where $F_G^{(i)}(\delta_i,\Tau_i)$ is the Hertz-Millis free energy for order with dynamical exponent $z_i$ defined explicitly in Ref \cite{m93}. While the free energy is described by the Gaussian part, the effect of the interactions is seen in the rescaling of the Gaussian parameters.

For a single QCP, the specific heat and thermal expansion behave differently in the quantum critical and Fermi liquid regimes, as tabulated in Ref. \cite{gr05}. In the multicritical case, in each of the four distinct regions of the paramagnetic phase in Fig. 3 the specific heat and thermal expansion are just the sum of the contributions from each individual QCP, which we find to leading order are unchanged by the interactions. In both two and three dimensions, in region I of the phase diagram in Fig. 3, where both ferro- and antiferromagnetic modes would expected be to quantum critical, then the strong temperature dependence of the ferromagnetic contribution dominates, and the presence of antiferromagnetic quantum criticality is subleading. In the other regions of the disordered phase, the observed quantities are the sum of the two terms. When the QCPs are separated, sufficiently close to each QCP the effects of the other QCP are not measurable and the thermodynamics is as would be expected from a single QCP.

The exception to this is in two dimensions in region I of the phase diagram in Fig. 3. In this case the temperature dependent renormalization of the antiferromagnetic tuning parameter is strong enough to push the antiferromagnetic mode out of the quantum critical regime, as the condition $R_2 < T$ will never be satisfied. This means we must use the Fermi liquid approximation ($R_2 > T$) in calculations of the antiferromagnetic contribution to physical quantities, but the correlation length is still dominated by temperature. However in this regime the ferromagnetic contributions dominate specific heat and thermal expansion and this effect is subleading.

%\section{Results}
\begin{figure}
\begin{center}
\includegraphics[scale=0.5]{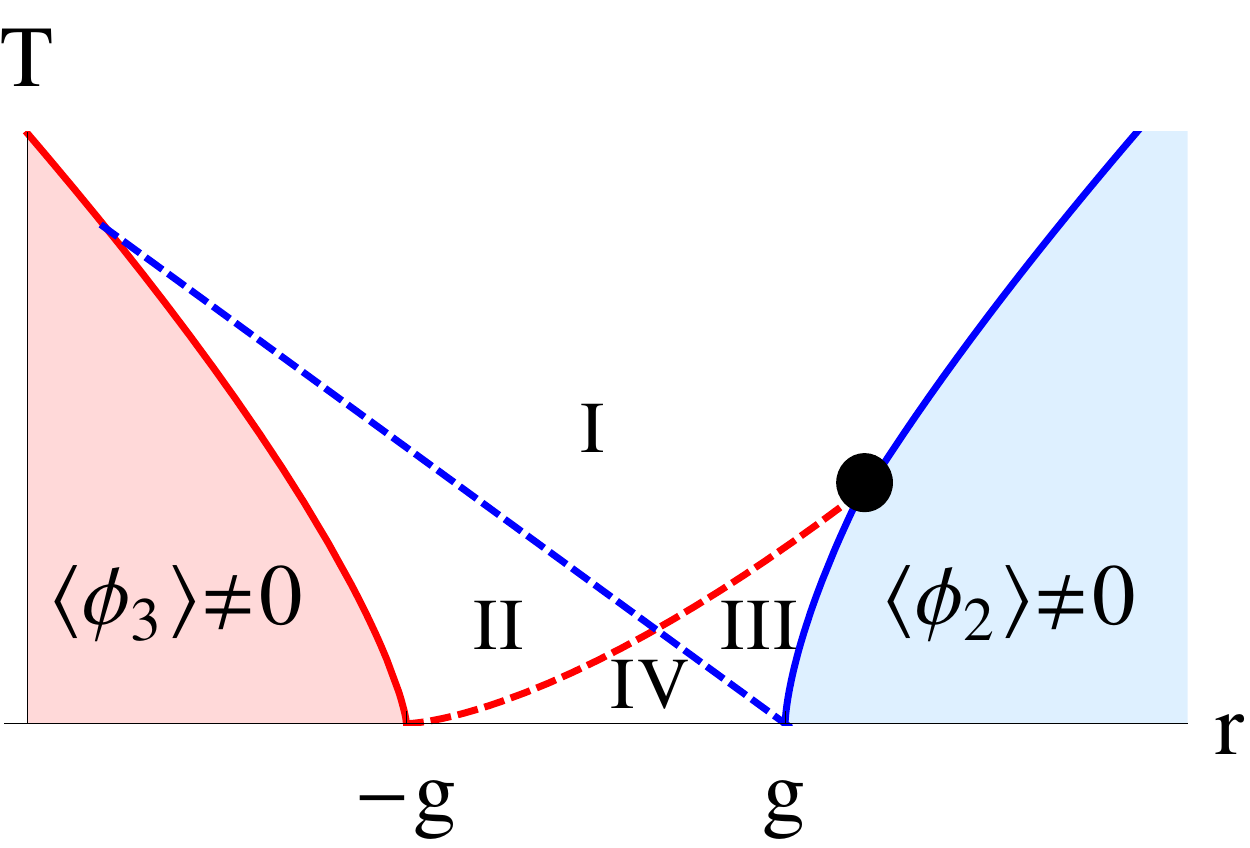}
\caption{Generic phase diagram of a quantum multicritical point, derived by setting $r_3=g+r$ and $r_2=g-r$. The regions in the paramagnetic regime have been identified. I and II are quantum critical regions where the ferromagnetic contributions dominate specific heat and thermal expansion. In I the ferromagnetic fluctuations control the antiferromagnetic correlation length and the boundary of the antiferromagnetic ordered phase. In II the antiferromagnetic correlation length is dominated by the tuning parameter. III is an antiferromagnetic quantum critical region, where the antiferromagnetic correlation length and boundary of the ordered phase are dominated by the antiferromagnetic fluctuations, and the ferromagnetic correlation length is dominated by the tuning parameter. IV is a Fermi liquid. The black dot indicates approximately where the antiferromagnetic phase boundary undergoes a crossover from a power law associated with antiferromagnetism to one normally associated with ferromagnetism. The subleading contributions depend on the dimensionality of space, as explained in the main text.}
\end{center}
\end{figure}

We now compare our theory with the existing experimental results in Nb$_{1-y}$Fe$_{2+y}$.
Near $y=0$ this material shows both 
ferro- and antiferromagnetic quantum critical points~\cite{tng10,hmb12}. 
There the measured specific heat shows a $-T \ln T$ dependence consistent with
the dominance of ferromagnetic fluctuations as we have shown above. The thermal exapansion has not been measured
but we predict it to show a $T^{1/3}$ dependence at low temperatures. In this work we have not calculated the resistivity because of the complex
interplay we anticipate between hot-spot/line scattering of the antiferomagnet~\cite{hr94} and the small angle scattering for the ferromagnetic
fluctuations.  The measured resistivity is $\Delta \rho \sim T^{3/2}$ which is consistent with a naive extension of our theory
with the antiferromagnetic fluctuations dominating~\cite{lrv07} because of their increased effectiveness in momentum relaxation when compared
to small $q$ scattering. A detailed analysis is left for future work. Similarly a more detailed doping-dependent study of the 
Ne\'el phase boundary $T_N(y)$ is necessary to compare with our predictions of a cross-over in power law. 

NbFe$_2$ is not unique in 
showing quantum multicriticality.  YbRh$_2$Si$_2$ orders antiferromagnetically at low temperatures 
but the specific heat and Gr\"{u}neisen parameter at low temperatures obey power laws as would be expected of a ferromagnet \cite{lrv07}. 
This could be the result of the presence of both ferro- and antiferromagnetic fluctuations~\cite{kbf05,abg14}. 
However, this material is usually thought to lie outside the Hertz-Millis
scenario for quantum criticality because of Kondo breakdown effects~\cite{nair_2012a}. 

%\section{Conclusions}
In summary we have analysed the interplay of two quantum critical points in an itinerant magnet in both two and three dimensions. Our main prediction is that if quantum critical fluctuations of both ferro- and antiferromagnetic order are present then the specific heat, thermal expansion and Gr\"{u}neisen parameter will have the temperature dependence associated with just the ferromagnetic modes. In addition, the correlation length of antiferromagnetic order will acquire the temperature dependence of the ferromagnetic correlation length, and this suppresses the boundary of the ordered phase (or region of applicability in two dimensions). We find that the boundary of the antiferromagnetic phase can undergo a crossover from its usual Hertz-Millis power law at low temperatures to the power law usually associated with a \emph{ferromagnetic} instability at higher temperatures. We find no difference between a quantum bicritical and a quantum tetracritical point, as under renormalization the system always flows to the Gaussian fixed point where the interactions are zero.

\begin{acknowledgments}
We would like to thank Sven Friedemann and Manuel Brando for sharing experimental data, and John Cleave for useful discussions. We acknowledge the funding support of EPSRC 
and grant EP/J016888/1. This research was also supported in part by the National Science Foundation under Grant No. NSF PHY11-25915.
\end{acknowledgments}

\bibliography{qmcbib4}% Produces the bibliography via BibTeX.

\end{document}